\documentclass{article}
\usepackage[utf8]{inputenc}
\usepackage{color}
\usepackage{float}
\usepackage{amsmath}
\usepackage{graphicx}
\usepackage{caption}
\usepackage{hyperref}

\makeatletter

\providecommand{\tabularnewline}{\\}

\usepackage{spconf}

\usepackage[utf8]{inputenc}

\title{The importance of spatial and spectral information in multiple speaker tracking}
\name{{Hanan Beit-On\textsuperscript{1}, Vladimir Tourbabin\textsuperscript{2} and Boaz Rafaely\textsuperscript{1}\thanks{This work was supported by Reality Labs Research @ Meta.}}}
\address{\textsuperscript{1} School of Electrical and Computer Engineering, Ben-Gurion University of the Negev \\
\textsuperscript{2} Reality Labs Research @ Meta
}

\@ifundefined{showcaptionsetup}{}{%
 \PassOptionsToPackage{caption=false}{subfig}}
\usepackage{subfig}
\makeatother

\begin{document}
\maketitle 
\begin{abstract}
Multi-speaker localization and tracking using microphone array recording
is of importance in a wide range of applications. One of the challenges
with multi-speaker tracking is to associate direction estimates with
the correct speaker. Most existing association approaches rely on
spatial or spectral information alone, leading to performance degradation
when one of these information channels is partially known or missing.
This paper studies a joint probability data association (JPDA)-based
method that facilitates association based on joint spatial-spectral
information. This is achieved by integrating speaker time-frequency
(TF) masks, estimated based on spectral information, in the association
probabilities calculation. An experimental study that tested the proposed
method on recordings from the LOCATA challenge demonstrates the enhanced
performance obtained by using joint spatial-spectral information in
the association.
\end{abstract}

\section{Introduction}

\label{sec:intro}

The localization and tracking of multiple speakers from microphone
array signals are essential for many applications, including speech
enhancement, robot audition, and spatial audio rendering. In a typical
scheme, a localization method first estimates the directions-of-arrival
(DOAs) of multiple concurrent speakers within short time segments.
Then, a tracking algorithm utilizes current and past DOA estimates
together with the speakers' dynamical model to infer a smooth DOA
trajectory for each speaker \cite{evers2020locata}. A number of multi-speaker
tracking algorithms have been published in the literature \cite{gehrig2006tracking,evers2020locata}.
However, these methods rely on spatial information alone. Consequently,
when spatial information is ambiguous or partial, e.g. for speakers
with similar DOA or following inactive speech segments (speakers moving
while quiet), ambiguity may arise in associating the DOA estimates
with the correct speaker. This ambiguity can lead to identity switches,
where the association algorithm consistently assigns the DOAs of one
speaker to another \cite{evers2020locata,Beit-On2022}. 

To address these limitations, recent multi-speaker localization and
tracking methods perform association based on spectral information
\cite{Beit-On2022,wang2019robust1,mack2022signal}. Speaker TF masks,
estimated by a monaural speech separation network based on spectral
information, are used to associate DOA estimates in the TF domain
to speakers. This approach has been shown to resolve the ambiguity
that arises when inactive speakers change positions \cite{Beit-On2022}.
However, the association in this approach is based on spectral information
alone, potentially leading to performance degradation when spectral
information is insufficient, especially for speakers of the same gender. 

Advance localization and tracking methods incorporate joint spatial-spectral
information in the association process to support correct association
when one of these information channels is incomplete \cite{subramanian2022deep,cao2021improved,shimada2022multi,lin2018jointly}.
The method in \cite{lin2018jointly} incorporates pitch estimates
as features in a filter that jointly tracks and separates multiple
speakers. In \cite{subramanian2022deep,cao2021improved,shimada2022multi},
deep neural network (DNN) based solutions were developed. In these
methods, convolutional layers are used as a feature extractor to produce
robust features (spectral and spatial) for detection and tracking.
However, these methods have mainly been studied in stationary scenes. 

This paper presents and investigates a multi-speaker tracking method
that uses joint spectral and spatial information for the association.
The proposed method incorporates speaker TF masks, estimated based
on spectral information, in a JPDA-based filter \cite{vo2015multitarget}.
Unlike the DNN-based methods, the modular design of the proposed method
enhances explainability, simplifies the analysis of the separation
and localization modules, and provides the ability to control and
analyze the integration of spatial and spectral information. These
advantages are reflected in the evaluation section that compares three
variants of the proposed method with spatial-based, spectral-based,
and joint spatial-spectral-based association on the LOCATA dataset,
demonstrating the clear advantage of joint spatial-spectral association
in challenging scenarios.

\section{System model}

Consider an $M$-element microphone array positioned in a reverberant
room with $Q$ speakers. The DOAs of the speakers relative to the
array are denoted by $\left\{ \Omega^{q}\right\} _{q=1}^{Q}$, where
$\Omega^{q}=\left(\theta^{q},\phi^{q}\right)$, with $\phi\in\left[-\pi,\pi\right)$
and $\theta\in\left[0,\pi\right]$ denoting the azimuth and elevation
angles, respectively. Let $\mathbf{p}_{t,f}\in\mathbf{C}^{M}$ denote
the STFT of the microphone signals, where $t$ and $f$ denote the
time and frequency indices, respectively. Assuming that speakers are
in the far-field of the array, $\mathbf{p}_{t,f}$ can be modeled
as:
\begin{equation}
\mathbf{p}_{t,f}=\sum_{q=1}^{Q}\mathbf{h}_{f}\left(\Omega_{t}^{q}\right)s_{t,f}^{q}+\mathbf{n}_{t,f}\label{eq:x_model-1}
\end{equation}
where $s_{t,f}^{q}$ is the STFT of the $q$'th speech signal at the
origin, $\mathbf{h}_{f}\left(\Omega_{t}^{q}\right)$ is the array
steering vector corresponding to the direction of the $q$'th speaker,
$\Omega_{t}^{q}$, which is now assumed to be a function of time,
and $\mathbf{n}_{t,f}$ is additive noise, which may represent reverberation,
sensor noise, and/or model error. The STFT window is assumed to be
sufficiently long compared to the length of the filters $\mathbf{h}_{f}$
in time such that the multiplicative transfer function (MTF) approximation
\cite{Avargel2007} holds. On the other hand, the STFT window is assumed
to be sufficiently short such that the DOAs are approximately constant
within time-frames. The latter assumption is a good approximation
when assuming typical values for the array and speakers' velocity,
and the time-frame duration employed for the audio signal processing
(frames typically shorter than 100 ms) \cite{tourbabin2016analysis}. 

We assume that the microphone signal is processed by a localization
algorithm that estimates the speaker DOAs in individual time-frequency
(TF) bins in the STFT domain, where at most one DOA is estimated at
a single TF bin, e.g. \cite{Nadiri2014,beit2020importance}. Let $\mathbf{z}_{t,f}=\left[\hat{\theta}_{t,f},\hat{\phi}_{t,f}\right]^{\text{T}}$
denote the DOA estimate from the $\left(t,f\right)$ bin and let $\mathbf{x}_{t}^{q}$
denote the state vector of the $q$'th speaker relative to the array.
The state vector includes the position of the speaker in Cartesian
or spherical coordinates, but it may also include other state variables
such as velocities, and accelerations. The goal of a multi-speaker
tracking algorithm is to sequentially estimate the speakers' trajectories,
$\mathbf{x}_{t}^{1},...,\mathbf{\mathbf{x}}_{t}^{Q}$ for $t>1$,
given the DOA estimate history, and the microphone signal up to time
$t$. 

\section{Multi-speaker tracking by spatial, spectral and joint speaker association}

Three widely used approaches for multi-target tracking are the JPDA,
multiple hypothesis tracking (MHT) and random finite set (RFS) \cite{vo2015multitarget}.
Here we introduce a JPDA-based filter for multi-speaker tracking.
The method serves as a platform to study the various association approaches:
spatial-based, spectral-based, and joint spatial-spectral-based. Nevertheless,
the insights gained from this study are relevant to other tracking
approaches. 

The original JPDA filter was designed to process multiple observations
per time step, while assuming that at most one observation can originate
from a single target. In the multi-speaker case, where targets are
speakers and observations are DOA estimates in the STFT domain, this
assumption does not hold since DOA estimates obtained at different
frequencies may originate from the same speaker. Therefore, we present
an adaptation of the JPDA filter, which processes a single estimate
at each cycle of the filter. Also, for simplicity, we do not consider
clutter, i.e. false alarms, in the formulation. 

\begin{figure}
\includegraphics[width=0.9\columnwidth]{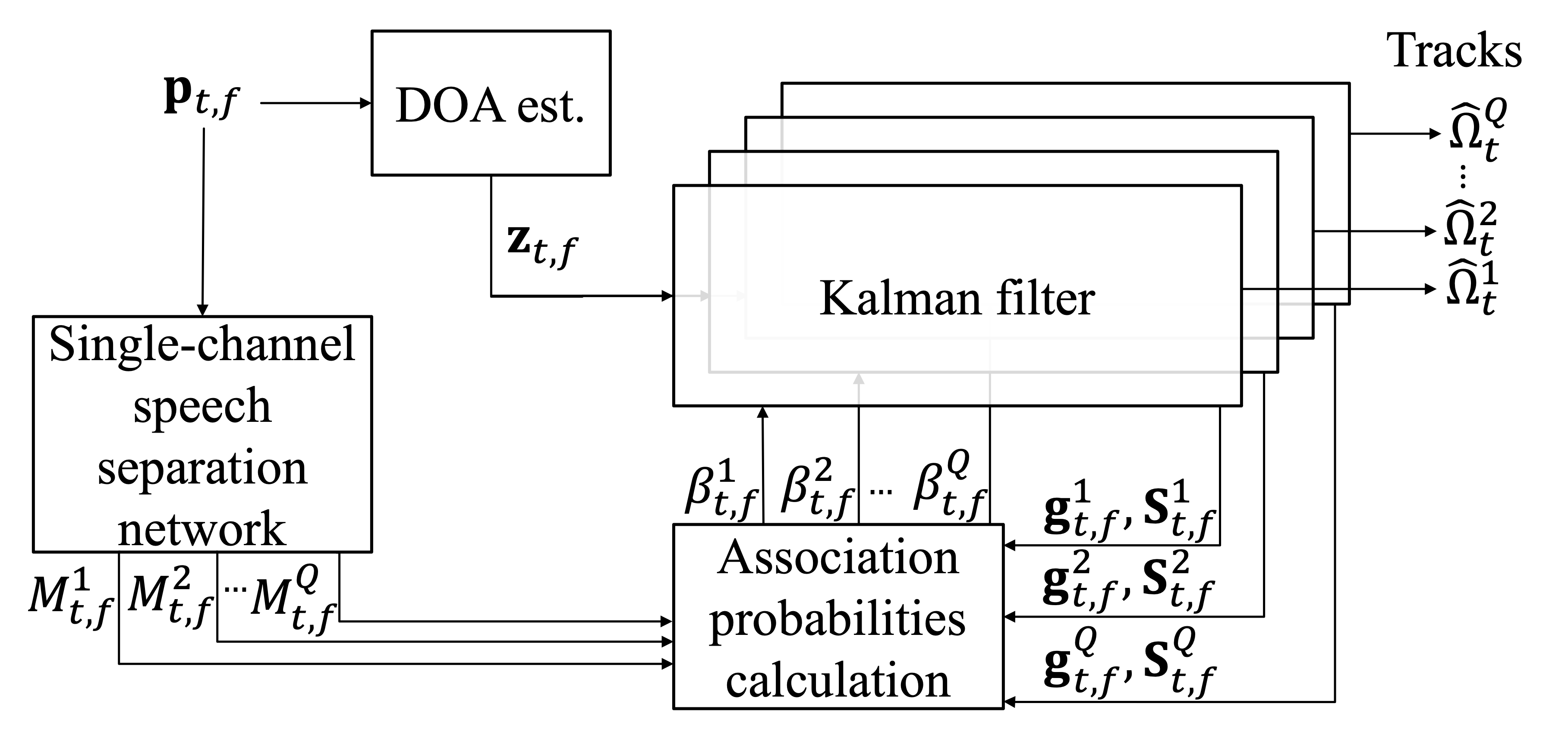}

\caption{\textcolor{red}{\label{fig:Block-diagram}}Block diagram of the proposed
method}
\vspace{-10bp}
\end{figure}

The JPDA runs Q Kalman filters in parallel (one for each speaker),
calculates DOA-estimate-to-speaker association probabilities and weighs
the innovation of each filter by its association probability. Let
$\mathbf{\hat{z}}_{t}^{q}$ denote the prediction of the $q$'th filter
at time $t$ and $\mathbf{g}_{t,f}^{q}=\mathbf{z}_{t,f}-\mathbf{\hat{z}}_{t}^{q}$
denote the innovation of the $q$'th filter due to the DOA estimate
$\mathbf{z}_{t,f}$, and let $\beta_{t,f}^{q}$ denote the association
probability, i.e. the posterior probability that $\mathbf{z}_{t,f}$
originated from speaker $q$. According to the JPDA approach, the
correction step of the $q$'th filter is obtained by using the weighted
innovation:

\begin{equation}
\tilde{\mathbf{g}}_{t,f}^{q}=\mathbf{g}_{t,f}^{q}\beta_{t,f}^{q}\label{eq:weighted innovation}
\end{equation}
As a result, the correction of the $q$'th filter is proportional
to its association probability $\beta_{t,f}^{q}$. Assuming that $\mathbf{g}_{t,f}^{q}\sim\mathcal{N}\left(\mathbf{0},\mathbf{S}_{t,f}^{q}\right)$,
the association probability with no clutter is given by \cite{vo2015multitarget}:
\begin{alignat}{1}
\beta_{t,f}^{q} & =c_{t,f}\mathcal{N}\left(\mathbf{g}_{t,f}^{q};\mathbf{0},\mathbf{S}_{t,f}^{q}\right)P_{t,f}^{q}\prod_{q'\ne q}\left(1-P_{t,f}^{q'}\right)\label{eq:ass_pr-1}
\end{alignat}
where $\mathcal{N}\left(\cdot;\mathbf{m},\mathbf{P}\right)$ is a
Gaussian probability density function (pdf) with mean $\mathbf{m}$
and covariance $\mathbf{P}$, $P_{t,f}^{q}$ is the detection probability
of speaker $q$, and $c_{t,f}$ is an appropriate normalization constant
ensuring $\sum_{q=1}^{Q}\beta_{t,f}^{q}=1$. As observed from (\ref{eq:ass_pr-1}),
$\beta_{t,f}^{q}$ increases for smaller innovations, i.e., for filters
whose predicted states are spatially close to the DOA estimate $\mathbf{z}_{t,f}$.
Without any prior information, speaker detection probabilities $P_{t,f}^{q}$
are typically selected to be uniform across speakers \cite{markovic2016multitarget,lin2018jointly}.
With this selection, the association with (\ref{eq:weighted innovation})
and (\ref{eq:ass_pr-1}) is determined based on spatial information
only, and association errors may occur when spatial information is
not sufficiently distinct, e.g. for spatially close speakers. To promote
correct association when spatial information is insufficient, we propose
to set the detection probabilities as:
\begin{equation}
{\color{red}{\color{black}P_{t,f}^{q}=M_{t,f}^{q}}}
\end{equation}
where $M_{t,f}^{q}\in\left[0,1\right]$ denote the TF mask of the
$q$'th speaker, which aims to separate the $q$'th speaker from the
signal. $M_{t,f}^{q}$ that is close to 1 indicates the dominance
of the $q$'th speaker within the $\left(t,f\right)$ bin, implying
high probability that the DOA estimate from this bin $\mathbf{z}_{t,f}$
originated from the $q$'th speaker. The speaker masks can be estimated
using speech separation networks, such as described in \cite{subakan2021attention},
which is provided with single microphone signal as input and optimized
during training using the true separated speech signals. With this
selection of the detection probabilities, the association probabilities
in (\ref{eq:ass_pr-1}) are based on both spectral and spatial information.
When the spectral information is not sufficiently distinct, the association
will be based mainly on the spatial expression and vice versa. Figure
\ref{fig:Block-diagram} depicts the diagram of the method. 
\vspace{-10bp}

\section{Experimental study}

\begin{figure*}[!t]
\subfloat[\label{fig:DOA-estimates-marked}DOA estimates; the color scale indicates
the mask value]{\includegraphics[width=1\linewidth]{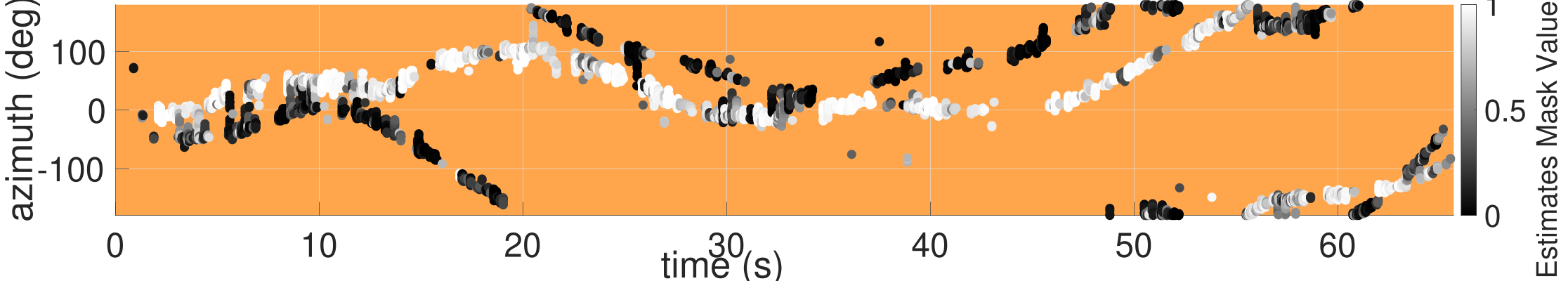}}\vspace{-10bp}

\subfloat[\label{fig:est-trajectories-spatial-1-1}Spatial speaker association]{\includegraphics[width=1\linewidth]{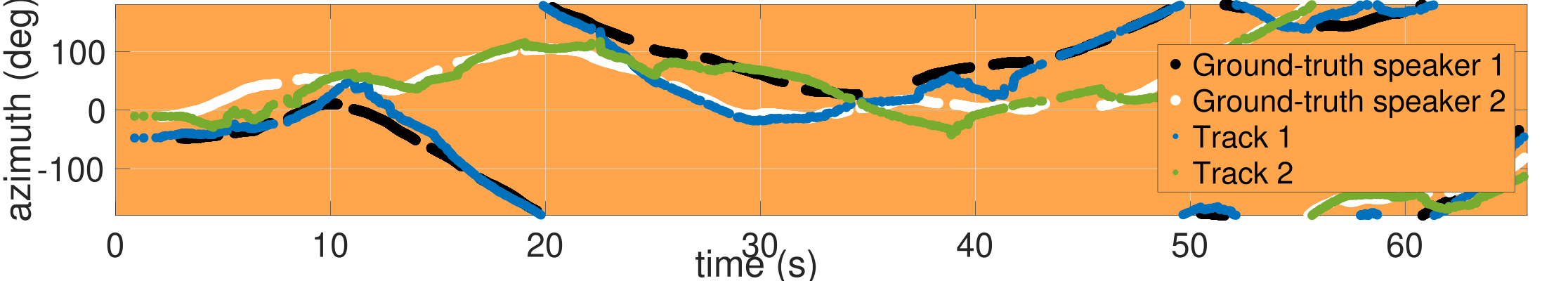}}\vspace{-10bp}

\subfloat[\label{fig:est-trajectories-spectral-1-1}Spectral speaker association]{\includegraphics[width=1\linewidth]{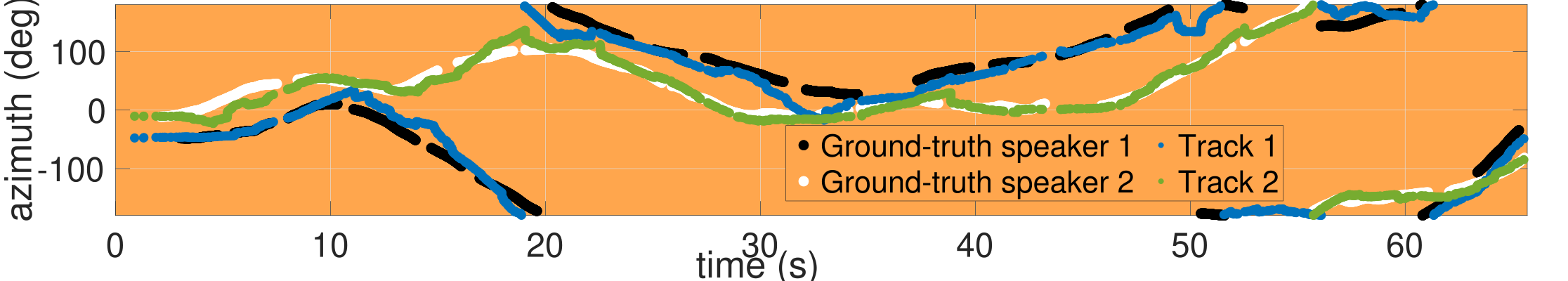}}\vspace{-10bp}

\subfloat[\label{fig:est-trajectories-joint-1-1}Joint spatial-spectral speaker
association.]{\includegraphics[width=1\linewidth]{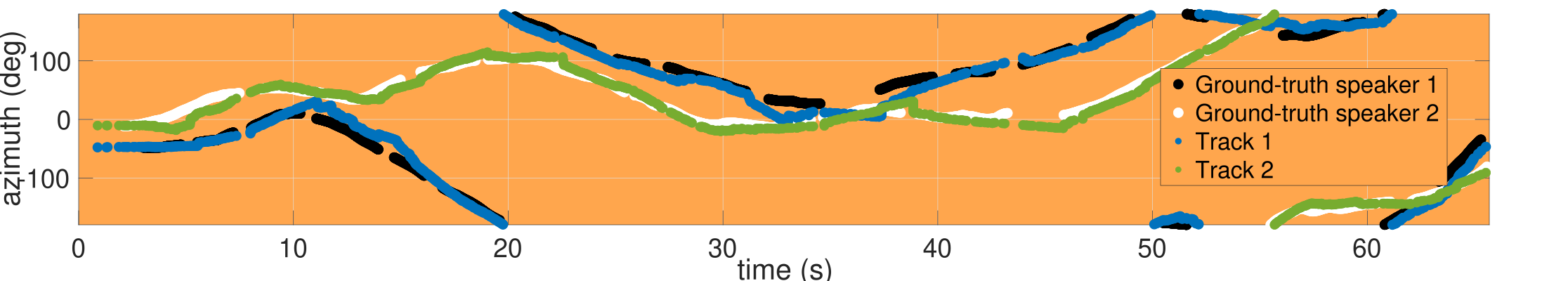}}\vspace{-10bp}

\caption{\label{fig:rec1task6}Results for recording 1 in task 6 of the LOCATA
challenge.}
\vspace{-10bp}
\end{figure*}
 This section compares spatial-based, spectral-based, and joint-based
association using the LOCATA data \cite{evers2020locata}.

\textbf{Setup -} The LOCATA challenge consisted of six tasks, with
multiple speakers involved in tasks 2, 4, and 6. Task 2 presented
static scenarios, while task 4 involved static arrays and moving speakers.
Task 6 involved both moving arrays and speakers. Tasks 4 and 6 each
comprised five recordings with two speakers, while task 2 included
13 recordings with 2-5 speakers. This experiment incorporated all
recordings form task 4 and 6 and recordings 10-13 from task 2, which
included two speakers, the scenario under study here. We chose to
evaluate the proposed method using recordings captured by the 12-microphone
array mounted on the Nao robot's head. 

\textbf{Methodology} - The microphone signals were downsampled from
48 kHz to 16 kHz and were transformed using the STFT with an FFT length
of 512 samples, a Hann window of 512 samples (32 ms), and a 50\% overlap.
The DOAs were estimated by the DPD test \cite{beit2020importance},
based on the Nao array steering vectors, which were available on the
challenge website. The DPD test was implemented with frequency smoothing
over $J=15$ frequencies and selected TF bins in the range 1-5 kHz,
with the threshold $\mathcal{TH}\left(f\right)$ chosen so that 6\%
of the TF bins at each frequency pass the test. The proposed JPDA-based
filter was implemented for azimuthal tracking. The wrapped Kalman
filter (WKF) \cite{traa2013wrapped} was employed that accounts for
the circular nature of the azimuth angle. The WKF filter incorporated
azimuth estimates as observations, i.e. $\mathbf{z}_{t,f}=\hat{\phi}_{t,f}$,
and included a state vector composed of azimuth and azimuthal velocity,
i.e, $\mathbf{x}_{t}=\left[\phi_{t},\dot{\phi}_{t}\right]^{T}$. The
separator employed was the SepFormer \cite{subakan2021attention},
known for its state-of-the-art performance, implemented with SpeechBrain
\cite{speechbrain1}, and trained on WHAMR! dataset \cite{maciejewski2020whamr},
which is a version of WSJ0-Mix dataset with environmental noise and
reverberation. The first microphone signal served as input to the
network, and the time-domain output signals were transformed by the
STFT to estimate the speaker ideal ratio masks (IRMs). Directly applying
the SepFormer, trained with approximately 4-second mixtures, to the
lengthy LOCATA mixtures led to suboptimal separation performance.
To overcome this mismatch, we partitioned the prolonged mixtures into
4-seconds segments and applied the separation process independently
to each segment and to the entire recording, producing accurate masks
for individual segments and less accurate masks for the entire recording.
Subsequently, we concatenated the short-segments masks to generate
masks for the longer recordings, utilizing the long-recording masks
to disentangle permutation in the concatenation process.

The proposed tracking algorithm was applied with three association
approaches: 1. \textbf{Spatial-based association (baseline approach):}
realized by setting the detection probabilities $P_{t,f}^{q}$ in
(\ref{eq:ass_pr-1}) to 0.5 for all speakers. 2. \textbf{Spectral-based
association:} realized by setting the Gaussian pdf term $\mathcal{N}\left(\mathbf{g}_{t,f}^{q};\mathbf{0},\mathbf{S}_{t,f}^{q}\right)$
in (\ref{eq:ass_pr-1}) to 1 for all speakers. 3. \textbf{Joint spatial-spectral
association:} realized by retaining both the Gaussian pdf term $\mathcal{N}\left(\mathbf{g}_{t,f}^{q};\mathbf{0},\mathbf{S}_{t,f}^{q}\right)$
and the detection probabilities $P_{t,f}^{q}$ in (\ref{eq:ass_pr-1}).

The initial state and its covariance were set to $\mathbf{\hat{x}}_{0}=\left[\hat{\phi}_{0},0\right]^{\text{T}}$
and $\mathbf{\hat{P}}_{0}=10^{-4}\cdot\mathbf{I}_{2}$, where $\hat{\phi}_{0}$
is an estimate obtained from clustering the DOA estimates from the
first $t_{0}$ seconds. $t_{0}$ was manually tuned for each recording
based on the time when the speakers were first active. For all recordings,
constant process and observation noise $\mathbf{Q}$ and $R$ were
used. We tuned $\mathbf{Q}$ and $R$ independently for each recording
to ensure optimal performance of the baseline method. The selected
parameters were kept unchanged for both the spectral and joint association
variants.

\textbf{Results} - Figure (\ref{fig:DOA-estimates-marked}) presents
the DOA estimates obtained by the DPD test, colored according to their
corresponding values in the second speaker's mask. For two speakers,
the IRM of the first speaker is the ones' complement of the second
speaker's IRM; therefore, low mask values in Fig. (\ref{fig:DOA-estimates-marked})
indicate the dominance of the first speaker. The figure shows that
in most cases, the estimates are close to the true directions and
that their color is close to that of the nearest speaker, implying
good localization and separation performance. Figure (\ref{fig:est-trajectories-spatial-1-1})
with spatial association shows identity switches around 25 and 35
seconds. These switches occur when speakers are in close proximity,
inhibiting accurate association based on spatial information alone.
Conversely, As can be seen from (\ref{fig:est-trajectories-spectral-1-1}),
spectral association prevents identity switches but result in reduced
accuracy. The association's dependence on the spectral information
alone leads to a greater sensitivity to the mask errors and, as a
result, drifts the tracking towards the other speaker. Finally, with
the proposed association, which combines spectral and spatial information,
there were no identity switches and the accuracy increased relative
to spectral association.

We use recording level mean absolute error (RLMAE) as a performance
measure. This error is calculated by assigning each track to the closest
speaker over the entire recording and calculates the mean absolute
error over the entire recording. The RLMAE significantly increases
for prolonged identity switches. Table \ref{tab:RLMAE-on-the} displays
the RLMAE of the proposed method variants. The results show that the
spatial association performance degrades significantly as the scenario
becomes more dynamic. In highly dynamic scenes, where speakers are
often proximate, spatial information becomes less reliable for association.
A more moderate performance degradation is observed for spectral association,
since spectral association is independent on dynamic conditions. The
joint spatial-spectral association achieves the smallest error, compensating
for the lack of spatial information with spectral information and
vice versa.

\begin{table}
\caption{RLMAE on the LOCATA challenge\label{tab:RLMAE-on-the}}

\begin{centering}
\begin{tabular}{|c|c|c|c|}
\hline 
 & Spatial & Spectral & Joint\tabularnewline
\hline 
\hline 
task 2 (rec. 10-13) & 3.74 & 7.97 & 3.21\tabularnewline
\hline 
task 4 & 8.06 & 10.01 & 8.52\tabularnewline
\hline 
task 6 & 17.64 & 14.17 & 10.95\tabularnewline
\hline 
\end{tabular}
\par\end{centering}
\vspace{-10bp}
\end{table}
\vspace{-10bp}

\section{Conclusions}

One of the main challenges in multi-speaker tracking is the correct
association of DOA estimates with the speakers. In this paper, we
proposed a multi-speaker tracking method that performs association
based on joint spatial-spectral information. The proposed method has
been shown to reduce identity switches and to increase tracking accuracy
compared to the spectral-based and spectral-based approaches. This
result shows the importance of combining these two information channels
in the association and suggests that incorporating such a combination
can improve other multi-speaker tracking methods.

\bibliographystyle{IEEEtran}
\bibliography{bib1}

\end{document}